\documentclass[11pt]{article}
\usepackage{jheppub}

\usepackage[usenames,dvipsnames,table]{xcolor}
\usepackage{graphicx,amsmath,amssymb,multirow,array,bm,mathrsfs}
\usepackage{epsf,amsfonts}
\usepackage[numbers,sort&compress]{natbib}
\usepackage{pdfpages}

\usepackage{slashed}

\newcommand{\be}{\begin{equation}}
\newcommand{\ee}{\end{equation}}
\newcommand{\bea}{\begin{eqnarray}}
\newcommand{\eea}{\end{eqnarray}}

\newcommand{\zbar}{{\bar z}}
\newcommand{\half}{{\frac{1}{2}}}
\newcommand{\TT}{[T^2]}
\newcommand{\OO}{{\cal O}}
\newcommand{\AAc}{{\cal A}}

\newcommand{\p}{\partial}

 \preprint{TCDMATH 19-10}

\title{Subleading Eikonal, AdS/CFT and Double Stress Tensors}
\author{Manuela Kulaxizi,}
\author{Gim Seng Ng and}
\author{Andrei Parnachev}
\affiliation{School of Mathematics and Hamilton Mathematics Institute,\\ Trinity College Dublin, Dublin 2, Ireland}
\emailAdd{manuela@maths.tcd.ie}
\emailAdd{gng@maths.tcd.ie}
\emailAdd{parnachev@maths.tcd.ie}

\abstract{
The  eikonal phase which determines the Regge limit of the 
gravitational scattering amplitude of a light particle off a heavy one in Minkowski spacetimes admits an expansion 
in the ratio of the Schwarzschild radius of the heavy particle to the impact parameter.
Such an eikonal phase in AdS spacetimes of any dimensionality has been computed to all orders and reduces to the corresponding Minkowski result
when both the impact parameter and the Schwarzschild radius are much smaller than the AdS radius.
The leading term in the AdS eikonal phase can be reproduced in the dual CFT by a single stress tensor
conformal block, but the subleading  term is a result of an infinite sum of the 
double stress tensor contributions.
We provide a closed form expression for the  OPE coefficients of the leading twist double stress tensors in four spacetime dimensions and perform
the sum to compute  the corresponding lightcone behavior of a heavy-heavy-light-light CFT correlator.
The resulting compact expression passes a few nontrivial independent checks.
In particular, it agrees with the subleading eikonal phase at large impact parameter.
}

\begin{document}
\maketitle

\section{Introduction and summary}
\label{sec:intro}
\subsection{Introduction}
The Shapiro time delay  and angle deflection of light are two of the four classic tests of general relativity \cite{Will:2014kxa}. 
A closely related object is the phase shift -- or equivalently the eikonal phase --
in gravitational high energy (Regge)  scattering (see e.g.  \cite{Giddings:2011xs} for a recent review).
Such a phase shift arises as a result of the eikonal resummation of graviton exchanges in 2-to-2 scattering amplitudes 
\cite{Dray:1984ha,tHooft:1987vrq,Amati:1987wq,Muzinich:1987in,Sundborg:1988tb}.
As explained in \cite{tHooft:1987vrq}, an alternative way of obtaining the phase shift for the 2-to-2 scattering of two light particles
involves studying the propagation of a lightlike particle in the background of a shock wave geometry (created by the second lightlike particle).
The null geodesic experiences a time delay as it travels through the shock wave;
the resulting phase shift can be computed as the product of this time delay and the lightcone momentum.

The leading eikonal approximation to the phase shift in gravity involves computing a single tree-level graviton exchange diagram, which then exponentiates.
To go beyond the eikonal approximation one needs to 
compute subleading diagrams.
This  has been performed in various regimes.
The situation which will be of particular interest to us involves high energy scattering of a light particle off a much heavier particle  (see e.g. \cite{Kabat:1992tb,DAppollonio:2010krb,Neill:2013wsa,Akhoury:2013yua,Bjerrum-Bohr:2014zsa,Bjerrum-Bohr:2016hpa,Luna:2016idw,Cachazo:2017jef,Bjerrum-Bohr:2018xdl,Cheung:2018wkq,Kosower:2018adc,Bern:2019nnu,KoemansCollado:2019ggb}).
In this case the subleading eikonal contribution comes from triangle-type diagrams
and is suppressed by the ratio of the Schwarzschild radius 
of the heavy particle to the impact parameter.

The AdS$_{d+1}$/CFT$_d$ correspondence \cite{Maldacena:1997re,Witten:1998qj,Gubser:1998bc} provides another angle on the subject of high energy scattering.
In  \cite{Cornalba:2006xk,Cornalba:2006xm,Cornalba:2007zb,Cornalba:2007fs,Costa:2012cb} this problem has been studied
for an AdS scalar field dual to a CFT operator $\OO$.
In the dual CFT language, the problem involves computing a four-point function in a certain kinematic limit (the Regge limit).
It receives contributions from  the stress tensor conformal block and conformal blocks of the $\OO \Box^n\p_\mu \p_\nu \OO$ double-trace operators. 
The Fourier transform of the amplitude produces the eikonal phase (a.k.a. the phase shift);
interestingly, it is only sensitive to the contribution from the stress tensor 
while double-trace operators decouple.
Using crossing symmetry one can also show that the phase shift encodes information on the anomalous dimensions of the $\OO \Box^n \p_{\mu_1} \ldots \p_{\mu_\ell} \OO$ operators,
which dominate in the cross channel.\footnote{ 
An interesting generalization of this story involves studying 2-to-2 Regge scattering of scalars and gravitons
(which corresponds to a four point function of two scalars and two stress tensors in the CFT language).
In \cite{Camanho:2014apa} it was shown that positivity of the time delay is equivalent to the absence of generic higher derivative corrections
to gravity. In the CFT language this translates into the statement that the imaginary part of the phase shift defined as a Fourier transform of the four point function, 
must be positive, and its positivity fixes the couplings of the stress tensor (which results in the ``$a=c$" condition in the superconformal case)
\cite{Kulaxizi:2017ixa,Costa:2017twz}; see also \cite{Afkhami-Jeddi:2016ntf,Li:2017lmh,Afkhami-Jeddi:2017rmx,Meltzer:2017rtf,Meltzer:2018tnm,Afkhami-Jeddi:2018apj,Belin:2019mnx,Kologlu:2019bco} for alternative derivations and generalizations. }

Consider now the AdS version of the high energy scattering of a light particle off a heavy one.
As in flat space, the amplitude is expected to exponentiate.
The leading  eikonal phase comes from the single-graviton exchange, 
while the first subleading correction  comes from the triangle Witten diagrams analogous to 
 the triangle diagrams in flat space\footnote{As explained in \cite{Kulaxizi:2018dxo}, the relevant Witten diagrams are semi-geodesic,
since the trajectory of the heavy particle's worldline does not fluctuate.}.
The eikonal expansion parameter $\mu$ is, roughly speaking, the Schwarzschild radius of the AdS-Schwarzschild black hole
measured in units of the AdS radius.
The amplitude also depends on the impact parameter $L$.

In  \cite{Kulaxizi:2018dxo} the AdS phase shift was computed to all orders in $\mu$ by studying null geodesics in a black hole background.
One can take the flat space limit of that result and compare with the corresponding eikonal phase of flat space amplitudes computed in the probe limit.
Consider Regge scattering in $D$-dimensional Minkowski spacetime, where a massless 
particle of energy $E$ scatters off a very heavy particle of mass $M$.
The leading $\delta^{(1)}$ and the subleading $\delta^{(2)}$  terms in the eikonal phase  can be found in  e.g. eqs. (3.7) and (3.9) of \cite{KoemansCollado:2019ggb}.
Taking the heavy-light limit 
\be
\label{hllimit}
m_2 =0, \;\;m_1 =M,\;\; s-m_1 = 2 M E
\ee
 the expressions become
\be
\label{leadingeik}
  \delta^{(1)} =   E b  \left(\frac{R_s}{b}\right)^{d-2} \; \frac{(d-1) \sqrt{\pi} \Gamma(\frac{d-3}{2})}{4 \Gamma(\frac{d}{2})}, \qquad 
    R_s^{d-2} = \frac{16 \pi G M}{(d-1) \Omega_{d-1}}
\ee
and
\be
\label{subleadingeik}
  \delta^{(2)} =   E  b  \left(\frac{R_s}{b}\right)^{2d-4} \; \frac{(4 d (d-2) +3) \sqrt{\pi} \Gamma(d-\frac{5}{2})}{16 \Gamma(d-1)}
\ee
where $b$ is the impact parameter, $d=D-1$ and $\Omega_{d-1}$ is the area of the $(d-1)$-dimensional sphere.
Restricting to the leading ($k=1$) and subleading  ($k=2$)  terms in eq. (2.33) of \cite{Kulaxizi:2018dxo} 
and identifying $\mu = R_s^{d-3}, \;\; L=b, \;\; \sqrt{-p^2} =E$ one   recovers
eqs.~(\ref{leadingeik}) and (\ref{subleadingeik}).

We expect the agreement above to persist to all orders in $\mu$.
In fact, we expect agreement beyond the flat space limit: one should be able to reproduce 
the result of  \cite{Kulaxizi:2018dxo}  by computing the corresponding AdS amplitudes (Witten diagrams)\footnote{It would be nice to verify this
by computing the corresponding amplitudes with graviton exchanges in AdS.}.
From the holographic point of view it is natural to identify the eikonal phase
  $\delta = -p \cdot \Delta x$  with the product of the boundary momentum $p$ and vector $\Delta x$, 
which specifies the time delay and angle deflection of the null geodesic  \cite{Kulaxizi:2018dxo}.
The argument goes as follows (see  \cite{Kulaxizi:2018dxo} for more details).
The Fourier transform of the four-point function in the dual CFT defines an on-shell amplitude in the 
dual gravity. 
The Regge limit implies that at large momentum $p$ the phase in the exponential simply picks up
the pole in the amplitude, which happens at the point $\Delta x$ where the corresponding null geodesic emerges at the boundary of AdS.

The gravity computation of the  eikonal phase  in AdS  \cite{Kulaxizi:2018dxo} has nontrivial implications for the dual CFT.
The four-point function of two light and two heavy operators  can in principle be computed by CFT methods.
If the gravity and the CFT computations agree, it would imply that the phase shift is an observable which does not distinguish 
a generic heavy state in the CFT from a {\it thermal} state (which is described in gravity by the black hole).
This was one of  the motivations of  \cite{Kulaxizi:2018dxo}, where
we considered  a $\langle \OO_H(\infty)   \OO_L(1)  \OO_L(z, \bar z) \OO_H(0)  \rangle$ correlator.
Here  $\OO_L$ is a light operator with conformal dimension $\Delta_L\sim O(1)$ and $\OO_H$ a heavy operator with $\Delta_H \sim O(C_T)$,
(the central charge $C_T$ of the conformal field theory is taken to be large).
The Regge limit corresponds to taking the two $\OO_L$  insertions close together.
In \cite{Kulaxizi:2018dxo} we showed that the expansion in powers of the Schwarzschild radius in gravity 
corresponds to the expansion of the correlator in powers of $\mu \sim \Delta_H/C_T$ in the CFT (note that $\mu$ remains fixed as both $\Delta_H$ and $C_T$ are taken to infinity).
Furthermore, we showed that the leading order $\mu$ result for the phase shift, as computed in gravity,
is exactly reproduced by the exchange of the stress-tensor in the ``T-channel" (in the limit $z, \bar z \to 1$ after an appropriate analytical continuation). 

Studying higher orders in $\mu$  in the CFT is more involved.
Given the gravitational result for the phase shift, in \cite{Karlsson:2019qfi} crossing symmetry was used to derive the anomalous dimensions and OPE coefficients of the 
double trace operators $[\OO_H \OO_L]_{n,l} \sim \OO_H \Box^n \p_{\mu_1} \ldots \p_{\mu_l} \OO_L $ contributing to the ``S-channel" ($z, \bar z \to 0$) expansion of the same correlator. The expressions obtained in \cite{Karlsson:2019qfi} precisely matched those independently computed in gravity up to next-to-leading order in $\mu$ in \cite{Kulaxizi:2018dxo}.
For this to happen, the thermalization of the heavy state in the holographic CFT was essential.

However, reproducing second and higher order terms in the gravitational phase shift {\it directly from the CFT}, remains a challenging endeavour.
Consider for example the T-channel expansion of the correlator at $O(\mu^2)$. Evaluation of the correlator in this case requires summing the contributions from the exchange of infinitely many double-trace operators built out of products of the stress tensor (henceforth referred to as double stress tensor operators\footnote{Things are slightly more subtle than that, but we still call them double stress tensor operators. The subtlety will be commented on later on.}).
In the standard analytic bootstrap approach, typically one studies kinematic limits where finitely many operators dominate in a particular channel. 
Harnessing the power of crossing symmetry then enables one to infer non-perturbative information about the spectrum and the OPE coefficients appearing in the other channel. 
In cases where both sides of the crossing equations  involve infinite sums, one often needs to perform the summation before taking any kinematics limit.
Unless one knows the OPE coefficients of these infinitely many operators, it is not clear how to proceed.
 
The $d=2$ case provides a solvable model, where these issues are very clearly illustrated.
In this case the infinite-dimensional Virasoro symmetry constrains the full stress tensor sector completely.
All multi-stress-tensor contributions can be summed into the Virasoro vacuum block, which can be computed at large $C_T$  \cite{Fitzpatrick:2014vua,Fitzpatrick:2015zha,Collier:2018exn}.
As explained in \cite{Kulaxizi:2018dxo}, to compute the phase shift  at $O(\mu^2)$, one needs to first sum over an infinite number of double stress tensor operators
of increasing spin.
This sum produces $\log^2 z$ and $\log z$ terms which after analytic continuation of $z$ around zero give rise to double and single poles at $z=1$. The leading pole is consistent with the exponentiation of the $\OO(\mu)$ result while the subleading pole encodes the value of the phase shift at order $\mu^2$.

In higher dimensions we have to work harder.
In this paper we will mostly consider the $d=4$ case, although we do not expect any conceptual differences in other dimensions.
In $d=4$, a set of OPE coefficients $C_{\OO\OO T^k}$ between two scalars and a multi-stress tensor have recently been computed  in \cite{Fitzpatrick:2019zqz}
for holographic CFTs\footnote{See also \cite{Li:2019tpf} for  generalizations involving the inclusion of matter in the bulk.}.
It has also been argued in  \cite{Fitzpatrick:2019zqz} that at each power of $k$, the OPE coefficients for the leading twist multi stress tensors are universal (independent of the higher derivative corrections to the bulk gravitational Lagrangian).
We conjecture a formula which fits all the OPE coefficients with the leading twist double stress tensors ($k=2$) listed in \cite{Fitzpatrick:2019zqz} (and is consistent with many 
more OPE coefficients which we computed by following the prescription of \cite{Fitzpatrick:2019zqz}).
Using this conjectured formula we managed to sum the contributions of all leading twist double-stress-tensor  operators
and hence obtain the $O(\mu^2)$ correlator in the lightcone limit $\zbar\rightarrow 1$.
Furthermore, the resulting compact and closed-form expression is used to extract the behavior of the $\langle \OO_H(\infty)   \OO_L(1)  \OO_L(z, \bar z) \OO_H(0)  \rangle$ correlator  both in the large impact parameter sector of the Regge limit and in the small $z$ lightcone region.
In the former case, an independent computation of the same correlator from the S-channel has been performed in \cite{Karlsson:2019qfi} and agrees perfectly with the  T-channel result of this paper.
In the latter case, the S-channel computation to $O(\mu^2)$ contained herein is new. 
We observe precise agreement in both cases.

These results yield an independent check of our proposal for the OPE coefficients of double stress tensor operators while providing strong evidence in favor of thermalization for generic heavy states in holographic CFTs.

\subsection{Summary of the results}
We will consider a four-dimensional unitary CFT with a large central charge $C_T\sim N^2$ , which admits an expansion 
around the mean free theory at $N=\infty$. 
An important consequence  is the existence of double trace operators in the spectrum of the theory -- they are required by crossing symmetry \cite{Komargodski:2012ek,Fitzpatrick:2012yx}.
We will also focus on CFTs whose spectrum of primary single-trace operators is characterised by an infinite gap $\Delta_{gap}$ separating operators of spin greater than two from the rest.

A notable class of double-trace operators in this context concerns those constructed from the stress tensor, with dimension $\Delta_{n,s}=4+2n+s$, spin $s$ and twist $\Delta_{n,s}-s=4+2n$
(to leading order in $1/C_T$):
\be\label{TTdef}
\TT_{n,s}\equiv T_{\mu\nu}\square^n \partial_{\mu_1}\ldots \partial_{\mu_{s-4}}T_{\rho\sigma}+\ldots\,.
\ee 
Here, the $\ldots$ include terms built out of descendants of stress tensor and terms required by symmetrization and tracelessness.
We are interested in extracting the OPE coefficient,
\be
C_{n,s}(\Delta) \equiv C_{\OO \OO \TT_{n,s}}
\ee 
between the double trace stress-tensor operators and two identical primary, scalar operators $\OO$ of dimension $\Delta$. Note that we have normalized the two-point functions of $\OO$ and all operators $[T^2]_{n,s}$ to have unit coefficient
(this fixes the form of the conformal blocks, discussed in the following section). 

Focusing on the lowest-twist double stress tensor OPE coefficient, i.e. the $n=0$ case, we propose that:
  \be
C_{0,s}(\Delta)= C_{\OO \OO \TT_{0,s}}
=\frac{160}{3}\frac{1}{C_T}
\frac{\Delta}{\Delta-2}a_{s}
\left[\Delta^2
+b_{s}\Delta+c_s
\right]
+O(1/C_T^2) \,,
 \ee  
for any $\Delta \ne 2$, where
\bea
\label{eq:bandc}
b_s&=&-1+\frac{36}{s(s+3)}+c_s \nonumber\\
c_s&=&\frac{288}{(s-2) s (s+3) (s+5)} 
\,.
\eea and
\be\label{eqas}
a^2_s=\frac{(s-2)  s(s+3) (s+5) (2 s+3) }{8 (s-3) (s-1) (s+1) (s+2) (s+4) (s+6)}\times\frac{\Gamma \left(s+2\right)^2}{ \Gamma (2 s+4)}\,.
\ee
It is very easy to check that  these equations are consistent with 
the OPE coefficients computed in \cite{Fitzpatrick:2019zqz} for a few double stress tensor operators
of low spin.
In fact, a slight generalization of the arguments of \cite{Fitzpatrick:2019zqz} allows one to prove
eq. (\ref{eqas}) -- this proof is presented in section \ref{app:geodesics}, where we also write down 
a generalization of (\ref{eqas})  to any $d$.
 
To verify (\ref{eq:bandc}) we perform two nontrivial independent checks.
As we review below,
 in the context of holographic CFTs, multi-stress-tensor operators play a central role in the determination of the heavy-heavy-light-light (HHLL) four-point function, 
\be\label{Gdefa}
G(z,\zbar) = \lim_{x_4\to\infty} x_4^{2\Delta_H}\langle\OO_H(x_4)\OO_L(1)\OO_L(z,\zbar)\OO_H(0)\rangle \,,
\ee
where $\OO_H$ denotes an operator whose conformal dimension $\Delta_H$ scales with $C_T$ , whereas $\OO_L$ denotes an operator whose dimension $\Delta_L$ is of order one (scales like $C_T^0$). 

In the limit  $\zbar \to 1$, the dominant contribution in the T-channel expansion $(\OO_L(z,\zbar)\rightarrow\OO_L(1))$ is due to the operators of lowest twist. Expanding further in powers of $\mu\sim{\Delta_H}/{C_T}$, one identifies the leading contribution coming from the identity operator at $\OO(\mu^0)$, the stress-tensor operator of twist two at $\OO(\mu)$ and the double-stress-tensor operators of lowest twist ($n=0$) at $\OO(\mu^2)$. Using the above OPE formula, one can sum over the relevant conformal blocks to 
the following lightcone behavior of the correlator at $\OO(\mu^2)$:
\be
\label{summaryresTfinal}
\begin{split}
\left.  G(z,\zbar) \right |_{\mu^2}  \,&\underset{\zbar\rightarrow 1}{\simeq} \frac{[(1-z)(1-\zbar)]^{-\Delta_L+2} }{(1-z)^2} \,\left( \frac{\Delta_L}{\Delta_L-2} \right) \frac{1}{28800} \times\\
&
\left[
(\Delta_L -4) (\Delta_L -3)
f_{3}(z)^2
+\frac{15}{7}(\Delta_L -8) f_2(z) f_4(z)
+\frac{40}{7}(\Delta_L +1) f_1(z) f_5(z)
 \right] 
 \end{split}
\ee 
 where
 \bea\label{eq:deff}
 f_{a}(z)\equiv (1-z)^a\,{}_2 F_1(a,a,2a,1-z) \,,
 \eea
and ``$\underset{\zbar\rightarrow 1}{\simeq}$'' implies equality up to subleading terms in $(1-\zbar)$.
It is impressive that the result is of such a  compact form. 
We shall discuss possible relations of the light-cone limit with ``eikonalization'' in 4d CFT in Section~\ref{sec:dis}.

The first non-trivial check of eq.~(\ref{summaryresTfinal}) involves taking the large impact parameter regime of the Regge limit. 
This indeed reproduces the corresponding expression obtained earlier  in \cite{Karlsson:2019qfi}. 
Another check involves taking the subsequent $z \to 0$ limit of the lightcone result eq.~(\ref{summaryresTfinal}).
This produces 
\be\label{lcTmutwo_intro}
\begin{split}
\left. G(z,\zbar)\right|_{\mu^2} \underset{\scriptscriptstyle{\bar z\to 1, \; z \to 0} }{\simeq}\,  (1-\zbar)^{2-\Delta_L}\, \frac{\Delta_L}{\Delta_L-2} \left\{ \frac{1}{32}\, \Delta_L(\Delta_L-1) \,  \log^2{z} + \frac{1}{16}  \left(3\Delta_L^2-7\Delta_L-1\right)\, \log{z} +\cdots   \right\}
\end{split}
\ee
which can be exactly matched to the respective limit of the correlator computed in the crossing channel,
provided  $\OO(\mu^2)$ anomalous dimensions of certain heavy-light double trace operators.
Fortunately, the results for these anomalous dimensions are available \cite{Kulaxizi:2018dxo}.
Again, we observe perfect agreement.

 \subsection{Outline}

The rest of the paper is organized as follows. 
In Section~\ref{sec:notation},  we set up  notations and write general expressions for the heavy-heavy-light-light correlator 
in the T- and S-channels.
In Section ~\ref{sec:Tchannel}, we use the conjectured OPE coefficients in the T-channel to perform the sum over the leading twist double stress tensor conformal  blocks, 
deriving an explicit $\OO(\mu^2)$ expression for the correlator in the lightcone limit $\bar z \to 1$. 
In Section~\ref{sec:Schannel}  we compute the subsequent $z\to 0$ behavior of the lightcone correlator using the S-channel data
and verify its agreement with the T-channel result.
We also show that the Regge limit of the correlator, obtained using the phase shift, agrees with the results of Section~\ref{sec:Tchannel}. Following that, in Section~\ref{app:geodesics}, we prove eq.~(\ref{eqas})
and generalise  it to any number of dimensions $d$. 
We discuss our results in Section~\ref{sec:dis}.
Appendix~\ref{app:2F1sq} proves a useful identity relating hypergeometric functions while Appendix~\ref{app:infinitesum} provides the details of the summations performed in Section~\ref{sec:Tchannel}. 
Appendix~\ref{app:2d} discusses the case of two-dimensional  CFT.

 \section{Heavy-heavy-light-light correlator in holographic CFTs}
 \label{sec:notation}
In this section, the crossing relations for a heavy-heavy-light-light correlator of pairwise identical scalars are reviewed. We consider large $N$ CFTs, with $N^2\sim C_T$ and $C_T$ the central charge, with a parametrically large gap $\Delta_{gap}$ in the spectrum of single trace operators with spin $J>2$. The object that we study is a four-point correlation function between two light scalar operators $\OO_L$, with scaling dimension $\Delta_L$ of order one, and two heavy scalar operators $\OO_H$, with scaling dimension $\Delta_H$ of  order $C_T$. The relevant correlation function is
\be\label{fourptfcn}
\langle \OO_H(x_4)\OO_L(x_3)\OO_L(x_2)\OO_H(x_1)\rangle = {\AAc(u,v)\over x_{14}^{2\Delta_H}x_{23}^{2\Delta_L}},
\ee
where $u,v$ are the cross-ratios defined as\footnote{Note the slightly non-standard definition of $(u,v)$ here.}
\be\label{CrossRatios}
\begin{split}
  u &= (1-z)(1-\zbar) = {x_{14}^2x_{23}^2\over x_{13}^2x_{24}^2}\\
  v &= z\zbar= {x_{12}^2x_{34}^2\over x_{13}^2x_{24}^2}\\
  \end{split}
\ee
and $x_{ij}=x_i-x_j$.
Conformal symmetry allows us to fix the positions of three out of four operators and focus on
\be\label{Gdef}
 G(z,\zbar) = \lim_{x_4\to\infty} x_4^{2\Delta_H}\langle\OO_H(x_4)\OO_L(1)\OO_L(z,\zbar)\OO_H(0)\rangle = {\AAc(z,\zbar)\over[ (1-z)(1-\zbar)]^{\Delta_L}} .
\ee
The four point function (\ref{Gdef}) can be expanded in the S-channel, $\OO_L(z,\zbar)\to\OO_H(0)$, as
\be\label{OPESChannel}
  G(z,\zbar)= (z\zbar)^{-{1\over 2}(\Delta_H+\Delta_L)}\sum_{\tau,\,\ell} P^{HL,HL}_{\tau,\ell}\, g_{\tau,\ell}^{\Delta_{HL},-\Delta_{HL}}(z,\zbar)\,,
\ee
where we defined
\be\label{Psdef}
P^{HL,HL}_{\tau,\ell}=\left(-{1\over 2}\right)^{\ell}\lambda_{\OO_H\OO_L\OO}\lambda_{\OO_L\OO_H\OO},
\ee
and $\Delta_{HL}=\Delta_H-\Delta_L$. Here $\lambda_{\OO_1\OO_2\OO_3}$  denote the relevant OPE coefficients and the sum runs over primaries $\OO$ of spin $\ell$ and twist $\tau\equiv \Delta-\ell$ with corresponding conformal blocks denoted by $g_{\tau,\ell}$.

Likewise, Eq.~(\ref{Gdef}) can  be expanded in the T-channel,  $\OO_L(z,\zbar)\to\OO_L(1)$, as follows
\be\label{OPETChannel}
 G(z,\zbar) = {1\over [(1-z)(1-\zbar)]^{\Delta}}\sum_{t,s} P^{HH,LL}_{t,s} g_{t,s}(1-z,1-\zbar),
\ee
where 
\be\label{Ptdef}
P^{HH,LL}_{t,s}=\left(-{1\over 2}\right)^{s}\, \lambda_{\OO_H\OO_H\OO}\lambda_{\OO_L\OO_L\OO}\,,
\ee
and the sum runs over primary operators $\OO$ of spin $s$ and twist $t$.

The equality of (\ref{OPESChannel}) and (\ref{OPETChannel}) constitutes an example of a crossing relation. In both channels the sum is over an infinite set of conformal blocks, each of which contains the contribution from a primary operator and all its descendants. 
Here, to distinguish between the generically different primary operators contributing in the T- and S-channel, we denoted their twists and spin by $(t,s)$ and $(\tau,\ell)$ respectively.

In this article we focus on the lightcone limit, $u\ll 1$ with $v=$ fixed, or equivalently, $\zbar\to 1$ and $z=$ fixed. We will further distinguish two cases: the first corresponds to the $u\ll v\ll 1$ regime of the lightcone limit, where $\zbar$ is sent to unity followed by $z\rightarrow 0$. The second is slightly more involved and makes contact with the Regge limit. It requires performing an analytic continuation of the lightcone limit result, by taking $z\rightarrow z e^{-2i\pi}$, followed by $z\rightarrow 1$. We will refer to the former case as the small $z$ or small $v$ regime of the lightcone limit and to the latter, as the large impact parameter region of the Regge limit. 

Finally, let us note that on general grounds we expect that the correlator can be expressed as a series expansion in the parameter $\mu$ defined as  
\be\label{MuDef}
\mu={{4\Gamma(d+2)}\over{(d-1)^{2}\Gamma(d/2)^{2}}}{{\Delta_{H}}\over{C_T}}, 
\ee
which is kept fixed as $C_T\to\infty$. Our conventions mostly follow those of \cite{Kulaxizi:2018dxo, Karlsson:2019qfi}.


\section {T-channel expansion in the lightcone limit}
\label{sec:Tchannel}
In this section we will focus on the T-channel expansion of the correlator. In particular, we will produce a closed form result to $\OO(\mu^2)$ in the lightcone limit, $u\ll 1$.

In the T-channel expansion, only the OPE coefficients may depend on the external operator's dimensions, we may thus write
\be
P^{(HH,LL)}_{t,s}=\sum_{k\ge 0} P^{(HH,LL);(k)}_{t,s} \mu^k\,,
\ee
which allows us to express the T-channel expansion as follows:
\be\label{GTmu}
G(z,\zbar) ={1\over [(1-z)(1-\zbar)]^{\Delta_L}}\sum_{k=0}^\infty \mu^k \left[\sum_{t,s}  P^{(HH,LL);(k)}_{t,s} g_{t,s}^{(HH,LL)}(1-z,1-\zbar)\right].
\ee 
Further taking the limit $u\ll 1$ results in 
\be\label{lightconea}
G(z,\zbar)\underset{u\ll1}{\simeq} u^{-\Delta_L}\sum_{k=0}^\infty \mu^k u^{\half t_{m}} (1-v)^{-{t_m\over 2}}\left[\sum_{s_m}  P^{(HH,LL);(k)}_{t_{m},s_m} f_{{t_{m}\over 2}+s_m}(v)\right],\,
\ee 
where we used the fact that the dominant contribution at each order in $\mu$ comes from operators of minimum twist denoted by $t_m$, and approximated the conformal blocks by \cite{Komargodski:2012ek,Fitzpatrick:2012yx}\footnote{Note that our notations here are slightly different from those in \cite{Komargodski:2012ek,Fitzpatrick:2012yx}.}
\be
\begin{split}
g_{t_m,s_m}^{(HH,LL)}(u,v)&\underset{u\ll1}{\simeq} u^{{t_m\over 2}} \, (1-v)^{-\frac{t_m}{2}}\,\,f_{{t_{m}\over 2}+s_m}(1-v),\\
f_{{t_{m}\over 2}+s_m}(v)&\equiv (1-v)^{{t_{m}\over 2}+s_m}~
{}_2 F_1\left[
\frac{t_{m}}{2}+s_{m},
\frac{t_{m}}{2}+s_{m},t_{m}+2s_{m},1-v
\right].
\end{split}
\ee
Note that the summation in brackets in (\ref{lightconea}) is over operators of the same minimum twist $t_m$ but varied spins $s_m$. For reasons of convenience we henceforth denote
\be\label{eq:defP2}
P^{(k)}_{t,s} \equiv P^{(HH,LL);(k)}_{t,s}\,.
\ee

To leading order in $\mu$, {\it i.e.}, $O(\mu^0)$, the dominant contribution to the correlator comes from the identity operator. Moving on to $O(\mu)$, single-trace operators of minimum twist dominate. These are conserved currents. In a generic holographic CFT without additional symmetries, we expect only one such conserved current, {\it i.e.}, the stress tensor operator. In this case, $t_m=d-2$ and $s_m=2$, while the OPE coefficients are fixed by the Ward identity to 
\be
P_{d-2,2}^{(1)}={\Delta_L\over 4} {\Gamma({d\over 2}+1)^2\over\Gamma(d+2)}\,.
\ee
The $\OO(\mu)$ contribution to the correlator is then
\be
\left. G(u,v)\right|_\mu \underset{u\ll1}{\simeq}  u^{-\Delta_L}\,u^{d-2\over 2}\,{\Delta_L\over 4} {\Gamma({d\over 2}+1)^2\over\Gamma(d+2)} \, (1-v)^2  {}_2 F_1\left[{d\over 2}+1,{d\over 2}+1,d+2,1-v \right]\,.
\ee
Further taking the limit, $v\rightarrow 0$, leads to:
\be\label{lcTmuone}
\left.G(u,v)\right|_\mu\underset{\scriptscriptstyle{u\ll v\ll 1}} {\simeq} \, -u^{-\Delta_L}\,u^{d-2\over 2} {\Delta_L\over 4}\left(2 \, H_{\tiny\frac{d}{2}}+\ln{v}\right)\,,
\ee
where $H_a$ denotes the $a$-th harmonic number.
 
Considering further terms quadratic in $\mu$, we focus on the contribution of double-trace operators built from the stress tensor. In higher dimensional CFTs there are three types of such operators, depending on whether none, one or two pairs of indices are contracted. The minimum twist ones necessarily belong to the first class, with twist $t_m=2(d-2)$.  There are infinitely many such operators of minimum twist, schematically denoted by $T_{\mu\nu}\partial_{\rho_1}\cdots\partial_{\rho_{s-4}}T_{\kappa\lambda}$, with spin ranging from $s_m=4$ all the way to infinity. Note however that only even spin operators contribute in the OPE of two identical operators, hence $s_m=4,6,8,\cdots$.
     
We are now ready to address the main objective of this section, which is to use the conjectured OPE coefficients to explicitly perform the summation over the infinite tower of operators contributing in the T-channel expansion at $\OO(\mu^2)$ in the limit $u\ll1$. In what follows we focus on holographic CFTs in $d=4$. Explicitly, we need to evaluate:
\be\label{totalsumTusmall}
\left. G(u,v)\right|_{\mu^2}\,  \underset{u\ll1}{\simeq}\,  u^{-\Delta+2} (1-v)^{2}\,\sum_{s=4,6,\cdots}^\infty P_{4,s}^{(2)} f_{2+s}(v)
\ee
where we dropped the subscript denoting minimum twist to simplify the notation.
To perform the sum we will use the conjectured OPE coefficients whose product reads\footnote{An analogous expression for $d=2$ is given in  Appendix \ref{app:2d}. In $d=2$, such an OPE formula can be proven by studying the heavy-heavy-light-light Virasoro vacuum block or by a direct construction of the double stress operators. Both of these methods are presented in Appendix \ref{app:2d}.}
\be\label{Ptot}
P_{4,s}^{(2)} =\frac{\Delta_L}{\Delta_L-2}a_{s}^2 \left[\Delta_L^2+b_{s}\Delta_L+c_s \right]\,,
\ee
with 
\bea\label{bcdef}
b_s&=&-1+\frac{36}{s(s+3)}+c_s\nonumber\\
c_s&=&\frac{288}{(s-2) s (s+3) (s+5)} 
 \nonumber\\
\eea 
and
\be
a^2_s=\frac{(s-2)  s(s+3) (s+5) (2 s+3) }{8 (s-3) (s-1) (s+1) (s+2) (s+4) (s+6)}\times\frac{\Gamma \left(s+2\right)^2}{ \Gamma (2 s+4)}\,.
\ee 

Next, we set $s\equiv 2m+4$, and split the sum in (\ref{totalsumTusmall}) into three parts:
\be\label{totalsumTusmallS}
\left. G(u,v)\right|_{\mu^2}\,  \underset{u\ll1}{\simeq}\,  u^{-\Delta_L} u^{2} \frac{\Delta_L}{\Delta_L-2} \left(S_a+S_b+S_c\right)\,,
\ee
where
\be\label{sumsa}
\begin{split}
S_a&\equiv \sum_{s=4,6,\cdots}^\infty a_s^2 (1-v)^{s} \, {}_2 F_1[2+s,2+s,8+2s,1-v ]\\
&=\sum_{m=0}^\infty q_2(m) (1-v)^{2m+4}\,{}_2F_1[6+2m,6+2m,12+4m,1-v]\,,
\end{split}
\ee
\be\label{sumsb}
\begin{split}
S_b&\equiv \sum_{s=4,6,\cdots}^\infty a_s^2 b_s\, (1-v)^s  {}_2F_1[2+s,2+s,4+2s,1-v]\\
&=-\sum_{m=0}^\infty q_2(m) (1-v)^{2m+4}\,{}_2F_1[6+2m,6+2m,12+4m,1-v]+\\
&\quad+\sum_{m=0}^\infty 36 \, q_1(m)\, (1-v)^{2m+4}\,{}_2F_1[6+2m,6+2m,12+4m,1-v]+\\
&\quad+\sum_{m=0}^\infty 288\, q_0(m)\, (1-v)^{2m+4}\,{}_2F_1[6+2m,6+2m,12+4m,1-v]  \,,\\
\end{split}
\ee and
\be\label{sumsc}
\begin{split}
S_c&\equiv \sum_{s=4,6,\cdots}^\infty  a_s^2 \,c_s\, (1-v)^s \, {}_2F_1[2+s,2+s,4+2s,1-v]=\\
&=288\sum_{m=0}^\infty q_0(m)\, (1-v)^{2m+4}\,{}_2F_1[6+2m,6+2m,12+4m,1-v]\,,
\end{split}
\ee
with
\be\label{coefqm}
\begin{split}
q_2(m)&\equiv \frac{\sqrt{\pi } 2^{-4 m-12} (m+1)^2 (m+2)^2 (2 m+7) (2 m+9) \Gamma (2 m+1)}{(m+3) (m+4) (m+5) \Gamma \left(2 m+\frac{11}{2}\right)} \\
q_1(m)&\equiv \frac{q_2(m)}{(2m+4) (2m+7)},\quad q_0(m)\equiv \frac{q_2(m)}{(2m+2)(2m+4)(2m+7)(2m+9)}\,.
\end{split}
\ee
It turns out that these three infinite sums can be performed with the help of the following identity for hypergeometric functions (for a proof see Appendix \ref{app:2F1sq}):
\be\label{eq:hyperid}
\begin{split}
&{}_2F_1[a,a,2a,w]\,{}_2F_1[b,b,2b,w]=\sum_{m=0}^\infty p[a,b,m] w^{2m} \,{}_2F_1[2m+a+b,2m+a+b,4m+2a+2b,w]\\
&p[a,b,m]=\frac{2^{-4 m} \Gamma \left(a+\frac{1}{2}\right) \Gamma \left(b+\frac{1}{2}\right) \Gamma \left(m+\frac{1}{2}\right) \Gamma (a+m) \Gamma (b+m) \Gamma \left(a+b+m-\frac{1}{2}\right) \Gamma (a+b+2 m)}{\sqrt{\pi } \Gamma (a) \Gamma (b) \Gamma (m+1) \Gamma \left(a+m+\frac{1}{2}\right) \Gamma \left(b+m+\frac{1}{2}\right) \Gamma (a+b+m) \Gamma \left(a+b+2 m-\frac{1}{2}\right)}\,,
\end{split}
\ee
by expressing the coefficients $q_i(m)$ in (\ref{coefqm}) as linear combinations of the $p[a,b,m]$ for certain values of $(a,b)$ (the interested reader may consult Appendix \ref{app:infinitesum} for more details). The result is:
\be\label{resTfinal}
\begin{split}
\left. G(z,\zbar)\right|_{\mu^2}&\underset{z\rightarrow 1}{\simeq}\frac{[(1-z)(1-\zbar)]^{-\Delta_L+2}}{(1-z)^2} \left( \frac{\Delta_L}{\Delta_L-2} \right)\frac{1}{28800} \times\\
&\times \left[
(\Delta_L -4) (\Delta_L -3)
f_{3}(z)^2
+\frac{15}{7}(\Delta_L -8) f_2(z) f_4(z)
+\frac{40}{7}(\Delta_L +1) f_1(z) f_5(z)
 \right] 
\end{split}
\ee
where we restored $u=(1-z)(1-\zbar)$, set 
 \bea\label{eq:deffa}
 f_{a}(z)\equiv (1-z)^a\,{}_2 F_1(a,a,2a,1-z) \,,
 \eea
and substituted $v\simeq z$ since we are working in the limit $\zbar\rightarrow 1$. This is the exact expression of the correlator in the lightcone limit at $\OO(\mu^2)$.

To check the validity of our result, we further determine the behavior of eq.(\ref{resTfinal}) in both the small $z$ region of the lightcone limit and the large impact parameter regime of the Regge limit. This is because in the aforementioned regimes it is possible to compute the same correlator from the S-channel expansion. 

Reaching the small $z$ lightcone limit starting from (\ref{resTfinal}), requires further expanding the expression obtained around $v\sim 0$, or equivalently, $z\ll 1$. 
This leads to:
\be\label{lcTmutwo}
\begin{split}
\left. G(z,\zbar)\right|_{\mu^2} \underset{\scriptscriptstyle{u\ll v\ll 1} }{\simeq}\,  (1-\zbar)^{2-\Delta_L}\, \frac{\Delta_L}{\Delta_L-2} \left[\frac{1}{32}\, \Delta_L(\Delta_L-1) \,  \log^2{z} + \frac{1}{16}  \left(3\Delta_L^2-7\Delta_L-1\right)\, \log{z} +\cdots   \right]\,.
\end{split}
\ee

On the other hand, to evaluate the correlator in the large impact parameter Regge limit starting from (\ref{resTfinal}), we first analytically continue $z$ around zero according to $z\rightarrow ze^{-2\pi i}$. We then set 
\be
z=1-\sigma\, e^\rho,\qquad \zbar=1-\sigma \,e^{-\rho}
\ee
and determine the behavior of (\ref{resTfinal}) for small values of $\sigma$, $\sigma\ll 1$. The result is:
\be\label{lcRegge}
\left.G(\sigma,\eta)\right|_{\mu^2}\underset{\scriptscriptstyle{\rho\gg 1,\,\sigma\ll 1}}{\simeq} \frac{1}{\sigma^{2\Delta_L}}\, \left[ -\frac{9\pi^2}{2}\,\frac{e^{-6\rho}}{\sigma^2 } \,\frac{\Delta_L (\Delta_L+1)(\Delta_L+2) }{\Delta_L-2}  +i\frac{35 \pi}{2} \,\frac{e^{-5\rho}}{\sigma}\,\frac{\Delta_L (\Delta_L+1)}{\Delta_L-2} +\cdots \right]\,.
\ee
Notice that the imaginary part behaves like $e^{-5\rho}/ \sigma$, in accordance with the expected behaviour for large impact parameters ($\rho\gg1$) \cite{Karlsson:2019qfi}.


\section{S-channel expansion in the lightcone limit}
\label{sec:Schannel}
In this section we will use the S-channel expansion of the correlator to determine its behaviour in both the small $z$ lightcone region and the large impact parameter region of the Regge limit at $\OO(\mu^2)$. We will further show that the result precisely matches eqs (\ref{lcTmutwo}) and (\ref{lcRegge}).

The starting point is the S-channel expansion of $G(z,\zbar)$,
\be\label{OPESChannela}
  G(z,\zbar)= (z\zbar)^{-{1\over 2}(\Delta_H+\Delta_L)}\sum_{\tau,\,{\ell}} P^{HL,HL}_{\tau,\ell}\, g_{\tau,\ell}^{\Delta_{HL},-\Delta_{HL}}(z,\zbar)\,,
\ee
with $P^{HL,HL}_{\tau,\ell}$ defined in (\ref{Psdef}). In a holographic CFT with an infinite gap in the spectrum of operators, the protagonists of the S-channel are double-trace primaries built out of one heavy and one light operator, schematically denoted by
\be\label{dtschannel}
\OO_H\square^n \partial_{\mu_1}\cdots\partial_{\mu_{\ell}}\OO_L\,.
\ee
It will be convenient in what follows to denote the dependence of the OPE coefficients on $(n,\ell)$ instead of $(\tau,\ell)$.

In the S-channel expansion corrections to the correlator arise from corrections to the conformal dimensions and OPE coefficients of the double-trace operators from their generalized free field theory values \cite{Fitzpatrick:2011dm}: 
\be
\tau=\Delta_H+\Delta_L+2n\ \,,
\ee and
\bea
&&P^{HL,HL;(0)}_{n,\ell}\nonumber\\
&=&\frac{\left(\Delta_L-1\right)_n \left(\Delta _H-1\right)_n \left(\Delta _L\right)_{\ell+n} \left(\Delta _H\right)_{\ell+n}}{\ell! n! \left(2+\ell\right)_n \left(n+\Delta _L+\Delta _H-3\right)_n \left(\ell+2 n+\Delta _L+\Delta _H-1\right)_\ell \left(\ell+n+\Delta _H+\Delta _L-2\right)_n}\,. \nonumber\\
\eea
Generally we expect that:
\be
\begin{split}
\tau=\Delta_H+\Delta_L+2n+&\gamma_{n,\ell}(\mu),\quad\text{with}
\quad \gamma_{n,\ell}(\mu)= \sum_{k=1}^\infty \mu^k\gamma_{n,\ell}^{(k)}\,, \\
\end{split}
\ee
and 
\be
P^{HL,HL}_{n,\ell}=\sum_{k=0}^\infty \mu^k P^{HL,HL;(k)}_{n,\ell}\,.\\
\ee
For convenience in what follows we drop the superscripts and set $P^{(k)}_{n,\ell}   \equiv P^{HL,HL;(k)}_{n,\ell}$.

In the regime where $\Delta_H$ is much larger than any other parameter of the system, the S-channel conformal blocks reduce to
\be\label{cblocks} 
g_{\tau,\ell}^{\Delta_{HL},-\Delta_{HL}}(z,\zbar) =   (z\zbar)^{{1\over 2}(\Delta_H+\Delta_L+\gamma_{n,\ell})}    \zbar^\ell \,,
\ee  
while the generalized free field OPE coefficients simplify considerably as well
\be\label{Pnotel}  
P^{(0)}_{n,\ell}  = {\ell^{\Delta_L-1} \over \Gamma(\Delta_L) \, . }  
\ee 

The region of interest here is $\Delta_H\gg \ell\gg 1$ and $n=0$. 
We expect from \cite{Kulaxizi:2018dxo} that the corrections to the conformal dimensions at large $\ell$ will behave as follows
\be
\gamma^{(k)}_{n,\ell} \approx \frac{\gamma^{(k)}_{n}}{\ell^{k}}.
\ee 
Since we are interested in $n=0$, we set $\gamma^{(k)}\equiv \gamma^{(k)}_0$ and $P^{(k)}_{\ell} \equiv P^{(k)}_{0,\ell}$.
In the large $\ell$ limit for $k\ge 2$ we further assume that
\be
P^{(k)}_{\ell}  \approx P^{(0)}_{\ell}   \frac{P^{(k)}}{\ell^{k}}\,,
\ee 
where we have denoted the $\ell$-independent coefficient as $P^{(k)}$.

The $\OO(\mu^0)$ term of the correlator can now be easily computed to yield
\be\label{munot}   
G(z,\zbar) |_{\mu^0} = \int_0^\infty \, d\ell\,  P^{(0)}_{\ell}  \zbar^\ell  = (-\log{\zbar})^{-\Delta_L}\underset{\zbar\rightarrow 1}{\simeq}\, {1\over (1-\zbar)^{\Delta_L}    }\,
\ee
where we replaced the summation with integration, valid for large $\ell$.   
Note that this is precisely the disconnected correlator in the  limit where $\zbar\to 1$ and $z\to 0$.

Moving on to the $\OO(\mu)$ term in the lightcone limit, we find:
\be\label{muone}   
G(z,\zbar) |_{\mu} = \int_0^\infty d\ell\,   P_{\ell}^{(0)} \left(  { P^{(1)}\over \ell} +  {\gamma^{(1)}\over 2 \ell}  \log{z}\right) \zbar^\ell  \underset{\zbar\rightarrow 1}{\simeq} \frac{1}{ (1-\zbar)^{\Delta_L-1}    }   \left(  \frac{P^{(1)}  }{\Delta_L-1} + \frac{\gamma^{(1)}}{2(\Delta_L-1) }  \log{z} \right) \,.    
\ee
The first order corrections to the OPE coefficients and conformal dimensions in the lightcone limit of the heavy-heavy-light-light correlator are not explicitly known. However we can easily compute them by matching (\ref{muone}) to the respective lightcone expansion from the T-channel, eq.(\ref{lcTmuone}). This yields,
\be\label{phgammaone}   
P^{(1)} ={3 \gamma^{(1)}\over2 },\qquad   \gamma^{(1)} =  -{  \Delta_L (\Delta_L-1) \over 2}    \,.
\ee

Finally, we consider the $\OO(\mu^2)$ terms. Expanding conformal blocks and OPE coefficients to quadratic order in $\mu$ in the regime $z\ll 1$ leads to:
\be
G(z,\zbar) |_{\mu^2} =  \int_0^\infty d\ell\,    P^{(0)}_{\ell}  \left[ {(\gamma^{(1)})^2\over 8 \ell^2}  \, \log^2{z}  + {\gamma^{(2)}+P^{(1)} \gamma^{(1)}\over 2 \ell^2} \,\log{z}
\right] \zbar^\ell    \,.
\ee
Integrating and keeping only the leading term as $\zbar\rightarrow 1$ yields:
\be
G(z,\zbar) |_{\mu^2}\underset{\scriptscriptstyle {z\ll \zbar \rightarrow 1}}{\simeq}\,  (1-\zbar)^{2-\Delta_L}\, \frac{\Delta_L}{\Delta_L-2} \left[ \frac{1}{32}\, \Delta_L(\Delta_L-1) \,  \log^2{z} + \frac{1}{16}  \left(3\Delta_L^2-7\Delta_L-1\right)\, \log{z} \right]\,,
\ee
where we used (\ref{Pnotel}), (\ref{phgammaone}) and the expression for $\gamma^{(2)}=- (\Delta_L -1) \Delta_L  (4 \Delta_L +1)/8$ in \cite{Kulaxizi:2018dxo}.
We find precise agreement with Eq.~(\ref{lcTmutwo}) obtained in the same limit from the T-channel expansion of the correlator. 

Let us now move on to the large impact parameter regime of the Regge limit. In this case, explicit results from the S-channel expansion already exist in the literature \cite{Karlsson:2019qfi}. In particular, eq (A.11) of \cite{Karlsson:2019qfi} is precisely the imaginary part of the correlator at $O(\mu^2)$ as computed from the S-channel expansion of the correlator for large impact parameter. Comparing with Eq.~(\ref{lcRegge}), we observe exact agreement.


\section{On  geodesics in AdS-Schwarzschild and $a_s^2$}
\label{app:geodesics}
In this section we argue that the $a_s^2$ term in the OPE coefficient comes purely from the square of the stress tensor block. This was termed ``eikonalization'' in the work of \cite{Fitzpatrick:2015qma} (see also \cite{Maxfield:2017rkn}). Here we will rehash the ideas and arguments of \cite{Fitzpatrick:2015qma,Maxfield:2017rkn} in our context.

Let us denote $\OO\equiv \OO_L$ and $\Delta\equiv \Delta_L$ in this section.
Consider the limit $1 \ll \Delta \ll C_T$. By the standard AdS/CFT dictionary, the (boundary) two-point function can be evaluated by the geodesic distance computed in the bulk geometry dual to the boundary heavy states:
\be
\lim_{\Delta \rightarrow \infty}  \langle \OO_H| \OO(x_3)\OO (x_4) |\OO_H\rangle
\approx e^{-\Delta \sigma_{reg}(x_3,x_4)}
\ee where the ends of the geodesic are anchored at $x_3$ and $x_4$. The geodesic distance $\sigma_{reg}$ is the regularized geodesic distance where the IR-divergence has been subtracted. This regularization procedure is done in pure AdS and has no effect on the $\mu$-dependent terms. We shall drop the subscript `reg' from now on. Note that $\sigma$ depends on $\Delta_H$ (but not $\Delta$) though we shall not display it explicitly so as not to clutter notations. Finally,  we shall refer to this $1 \ll \Delta \ll C_T$ limit as the `geodesic limit'.

We will extract the $\mu$-dependent terms following the discussion in Section 3.2.3 of \cite{Fitzpatrick:2019zqz} for (spherical) AdS black holes. Studying the geodesic limit does not allow us to determine the exact lowest-twist double-stress tensor OPE coefficients. However, it enables us to compute the correct $a_s^2$ part of the OPE. This is due to the fact that the relevant OPE coefficient is rational function of $\Delta$ which in the limit $\Delta \gg 1$  gives 
\be
\lim_{\Delta \gg 1} P_{4,s}^{(2)}  = a_s^2 \Delta^2.
\ee 
Here  $P^{(2)}_{4,s}$  denotes the product of the T-channel OPE coefficients defined in Eq.~(\ref{Psdef}) and Eq.~(\ref{eq:defP2}).

In fact, we will not need to compute the geodesic distance $\sigma$ explicitly but will make use of the fact that it can be expanded as a power series in $\mu$:
\be\label{sigmaseries}
\sigma=\sum \mu^k \sigma_{(k)}\,.
\ee 
Substituting this expansion into the equation above and focusing on the $\mu^2$ term, we obtain
\bea
&&\lim_{\Delta \rightarrow \infty}  \langle \OO_H| \OO(x_3)\OO (x_4) |\OO_H\rangle\nonumber\\
&\approx& e^{-\Delta \sum \mu^k \sigma_{(k)}}
\approx e^{-\Delta \sigma_{(0)}}
\left[
1-\Delta \mu \sigma_{(1)}
+\left(
\half \sigma_{(1)}^2  \Delta^2
+O(\Delta)\right)\mu^2
+O(\mu^3)
\right].
\eea  
Note that, as discussed in \cite{Fitzpatrick:2015qma}, the $\mu^2 \Delta^2$ term is exactly equal to half of the square of the term linear in $\mu$.

Now, since the  $\OO(\mu)$ term $\sigma_{(1)}$ reproduces the full stress tensor block (c.f. \cite{Fitzpatrick:2019zqz}), the $\mu^2 \Delta^2$ term should be equal to one-half of the the square of the stress tensor block times the OPE coefficient.
Note that this is true for general $x_3$ and $x_4$ or cross-ratios $u$ and $v$. In particular, in the light-cone limit, the stress-tensor block with its associated OPE coefficient is:
\be
-\sigma_{(1)}=\frac{1}{120}\times (1-v)^2 ~{}_2F_1\left(3,3,6,1-v\right)
\ee and so 
\be\label{sigmasquare}
\half \sigma_{(1)}^2 = \half \left[\frac{1}{120}\times (1-v)^2 ~{}_2F_1\left(3,3,6,1-v\right)\right]^2.
\ee 
Eq.~(\ref{sigmasquare}) can now be decomposed into a sum over an infinite number of lowest-twist double-stress tensor conformal blocks as explained in detail in (\ref{eq:Tsqcontribution}), with the help of the identity  (\ref{eq:hyperid}).  As a result one proves that indeed $a_s^2$ is given as in Eq.~(\ref{eqas}).

Let us finish this section with a few comments on possible generalizations of the above arguments:
\begin{enumerate}
\item Clearly this section's result is not confined to $d=4$. One can obtain $a_s^2$ by the same method in any dimension. By squaring the stress-tensor block with its associated OPE coefficient, the $\mu^2 \Delta^2$ term in any dimension is given by
\be
\frac{1}{2}
\left[
 \frac{\Gamma \left(\frac{d+2}{2}\right)^2}{4 \Gamma (d+2)}
\times (1-v)^2
\times {}_2F_1\left(\frac{d+2}{2},\frac{d+2}{2};d+2;1-v\right)
\right]^2
\ee which when decomposed into the light-cone blocks yields the coefficient
\be
a_s^2=\frac{4^{2 d+s-2} \Gamma \left(\frac{d+3}{2}\right)^2 \Gamma \left(\frac{s-3}{2}\right) \Gamma \left(d+\frac{s-1}{2}\right) \Gamma \left(\frac{1}{2} (d+s-2)\right)^4}{\pi ^2 \Gamma \left(\frac{d}{2}+1\right)^2 \Gamma \left(\frac{s}{2}-1\right) \Gamma \left(d+\frac{s}{2}\right) \Gamma (2 d+2 s-5)} \,.
\ee

\item So far we have not discussed how higher order corrections to the geodesic distance (\ref{sigmaseries}) impact the correlator. In principle, one can compute $\sigma_{(2)}$ from an expansion in $\mu$ of the bulk geodesic and use it to determine the $\mu^2 \Delta$ term in the correlator. However, the situation is more complicated as $\mu^2 \Delta$ terms may also originate from $1\over\Delta$-corrections to the geodesic approximation \cite{Maxfield:2017rkn}. 

\item Finally, as was first discussed in \cite{Fitzpatrick:2015qma}, at order $\mu^k$, the leading $\Delta$-term ({\it i.e.}, $\Delta^k$ term) is given by $(1/k!) \sigma_{(1)}^k$ exactly.  There exists a similar decomposition of higher-powers of the stress tensor block into sums over triple and higher-stress-tensor blocks. It would be interesting to explore such structures.

\end{enumerate}


\section{Discussion}
\label{sec:dis}
We conclude this paper with a few generalizations, discussions and speculations.

Let us start by pointing out that the $\bar z\rightarrow 1$ limit of the correlator computed here in four dimensions has a  structure very similar to the Virasoro vacuum block in two dimensions\footnote{In Appendix \ref{app:2d} we discuss the $d=2$ case in detail.}. The two dimensional case was discussed around eq.~(4.12) of \cite{Kulaxizi:2018dxo}, where it was noticed that 
the $\OO(\mu^2)$ term in the Virasoro vacuum is a sum of two terms, each of which is  a product of two hypergeometric functions. 
It is intriguing that we find a similar structure in the lightcone limit of the heavy-heavy-light-light correlator in four dimensions. 

This leads us to the following conjecture for the lightcone behavior of the heavy-heavy-light-light correlator in arbitrary even dimensions: 
\be
G(z)|_{\mu^2}\sim  \sum_{i=1}^{(d+2)/2} A^d_i(\Delta_L) f_{i}(z) f_{d+2-i}(z)\,.
\ee  
where $f_i$ is the hypergeometric function defined in eq.~(\ref{eq:deff}).
It would be nice to verify this expression and evaluate the coefficients $A^d_i$ in any $d$.

In two dimensions the coefficients in the expansion of the Virasoro vacuum block can be determined
by a  set of diagrammatic rules \cite{Fitzpatrick:2015foa}. It would be interesting to explore the possibility of deriving similar rules  for the lightcone limit of the four-dimensional correlator.

A related direction, as mentioned in the introduction, involves the study of the subleading eikonal terms in AdS, given by corresponding Witten diagrams. 
It is plausible that the diagrammatic rules set forth in \cite{Fitzpatrick:2015foa} for large-$C_T$ two-dimensional CFTs can be reformulated at the level of AdS Witten diagrammatic rules.
If this were true, one could hope to reverse engineer the  CFT diagrammatic rules and generalize \cite{Fitzpatrick:2015foa} to higher dimensions.

So far we have been discussing $O(\mu^2)$ terms. 
In two dimensions, it is clear from eq (4.12) of \cite{Kulaxizi:2018dxo}, that the Virasoro vacuum block structure retains similar features at order $O(\mu^k)$ --
here one finds a sum where each term is a product  of $k$ hypergeometric functions.
It would be interesting to generalize the OPE formula and the summation we did in this paper to higher orders in $\mu$ to explore whether the analogy with two dimensions persists.
If this were true, one could potentially hope to build a closed-form expression for the four dimensional heavy-heavy-light-light correlator in the lightcone limit.

It is natural to inquire whether a direct computation of the OPE coefficients can be done using purely  CFT techniques. As we  explicitly show in Appendix \ref{app:2d}, this is indeed possible in two dimensions. 
A direct construction of the double-stress-tensors and the computation of the OPE coefficients\footnote{For some work related to the explicit construction of double-trace operators, see \cite{Penedones:2010ue,Fitzpatrick:2011dm,Minahan:2014usa,Arutyunov:2000ku}.} appears challenging in higher dimensions, yet the results of this paper indicate that it may be possible in the lightcone limit.

Finally, recall that in two dimensional  CFTs, the Virasoro algebra implies the existence of infinitely many commuting conserved charges, called the KdV charges \cite{Bazhanov:1994ft}. This integrable structure is expected to play an important role in understanding thermalization in holographic two-dimensional CFTs beyond leading-$C_T$. This in turn will allow for a more thorough and systematic study of the quantum corrections of BTZ black holes.\footnote{See \cite{deBoer:2016bov,Maloney:2018hdg,Maloney:2018yrz,Dymarsky:2018lhf,Dymarsky:2018iwx,Dymarsky:2019etq,Datta:2019jeo,Brehm:2019fyy} for recent work on understanding generalized Gibbs ensemble in this context.}   
Generalizing this story to higher dimensions, one may hope to discover  hidden integrable structures in holographic  CFTs. Could this be related to various integrable structures\footnote{For an inexhaustive list of reviews and recent works on this subject, we refer the readers to \cite{Leigh:2014dja,Petropoulos:2015fba,Petkou:2015fvh,Klemm:2015uba,Gath:2015nxa,Chervonyi:2015ima,Frolov:2017kze,Bena:2017upb} and the references within.} of general relativity?

\section*{Acknowledgements}
We thank Sergey Frolov, Riccardo Gonzo, Robin Karlsson, Arthur Lipstein, Raul Pereira and Petar Tadic for discussions. G. N. is supported by Simons Foundation Grant to HMI under the program ``Targeted Grants to Institutes''. 
The work of A.P. is supported in part by an Irish Research Council Consolidator Laureate Award.

\appendix


\section{Identities for a product of hypergeometric functions}
\label{app:2F1sq}
We would like to show that the following identity 
\be\label{eq:hyperida}
{}_2F_1[a,a,2a,w]\,{}_2F_1[b,b,2b,w]=\sum_{m=0}^\infty p[a,b,m] w^{2m} \,{}_2F_1[2m+a+b,2m+a+b,4m+2a+2b,w]\,
\ee
with
\bea
&&p[a,b,m]
\nonumber\\
&\equiv&\frac{2^{-4 m} \Gamma \left(a+\frac{1}{2}\right) \Gamma \left(b+\frac{1}{2}\right) \Gamma \left(m+\frac{1}{2}\right) \Gamma (a+m) \Gamma (b+m) \Gamma \left(a+b+m-\frac{1}{2}\right) \Gamma (a+b+2 m)}{\sqrt{\pi } \Gamma (a) \Gamma (b) \Gamma (m+1) \Gamma \left(a+m+\frac{1}{2}\right) \Gamma \left(b+m+\frac{1}{2}\right) \Gamma (a+b+m) \Gamma \left(a+b+2 m-\frac{1}{2}\right)}\,\nonumber\\
\eea
is true.

This is fairly simple once one realizes that the identity above is actually a special case of a more general one, known as the multiplication formula,
\be\label{mulfor}
\begin{split}
&{}_2F_1[a_1,a_2,a_3,w]\,{}_2F_1[b_1,b_2,b_3,w]=\sum_{n=0}^\infty \left\{\frac{(a_1)_n(a_2)_n (b_3)_n}{n!\,(a_3)_n(a_3+b_3+n-1)_n} \right.\\ 
&\left. \,{}_3F_2[\{b_1,1-a_3-n,-n\},\{b_3,1-a_1-n\},1]\,{}_3F_2[\{b_2,1-a_3-n,-n\},\{b_3,1-a_2,-n\},1] \right\}   \times\\
&\qquad\qquad\qquad\qquad\qquad\qquad\quad\times w^n\,{}_2F_1[a_1+b_1+n,a_2+b_2+n,a_3+b_3+2n,w]\,,
\end{split}
\ee
and was proven in Sec~10 of \cite{Burchnall48}.
To arrive at (\ref{eq:hyperida}) one simply needs to set $a_1=a_2=a,\,a_3=2 a$ and $b_1=b_2=b,\,b_3=2 b$ and observe that the relevant ${}_3F_2$  simplify considerably due to the identity
\be\label{f32id}
{}_3F_2\left[\left\{c_1,c_2,c_3\right\},\left\{2 c_3,\frac{1+c_1+c_2}{2}\right\},1\right]= \frac{\sqrt{\pi}\Gamma[c_3 + \half] \Gamma[\frac{c_1 + c_2 + 1}{2}] \Gamma[c_3 + \frac{1 - c_2 - c_3}{2}]}{\Gamma[\frac{c_1 + 1}{2}] \Gamma[\frac{c_2 + 1}{2}] \Gamma[c_3 + \frac{1 - c_1}{2}] \Gamma[c_3 + \frac{1 - c_2}{2}]} 
\ee
valid as long as $2 c_3-c_1-c_2>-1$.
This allows one to show firstly that all the coefficients of odd natural numbers $n$ in (\ref{mulfor}) vanish identically, and prove secondly that all the ones for even natural numbers $n=2m$ reduce to
\be
\frac{(a)_{2m}(a)_{2m} (2 b)_{2m}\,\left({}_3F_2[\{b,1-2 a-2m,-2m\},\{2b,1-a-2m\},1]\right)^2}{(2m)!\,(2 a)_{2m}(2 a+2b+2m-1)_{2m}}   =p[a,b,m]\,.
\ee


\section{On performing the infinite sums}
\label{app:infinitesum}
In this appendix we will present some further details regarding the summations performed in Sec.~\ref{sec:Tchannel}. The starting point is Eq.~(\ref{totalsumTusmallS}) expressed in terms of the $S_a,S_b$ and $S_c$ as defined in (\ref{sumsa})-(\ref{sumsc}):
\be
\left. G(u,v)\right|_{\mu^2}\,  \underset{u\ll1}{\simeq}\,  u^{-\Delta_L} u^{2} \,(S_a+S_b+S_c)\,.
\ee
In what follows we will set $v=1-w$ for convenience. Let us consider first the term proportional to $\Delta^2$, {\it i.e.}, $S_a$. In this case, it is easy to see that
\be
q_2(m)=\frac{1}{2}\times\frac{1}{120^2}\,\, p[3,3,m]
\ee
with $p[a,b,m]$ as defined in (\ref{eq:hyperida}). This allows us to immediately write
\be \label{eq:Tsqcontribution}
S_a=\half \times{1\over 120^2}\times \, \left(w^2\,{}_2F_1[3,3,6,w]\right)^2
\ee
Next we wish to compute the term proportional to $\Delta$ in (\ref{Ptot}), {\it i.e.} $S_b$. As seen from (\ref{bcdef}) and (\ref{sumsb}), this term can decomposed into three separate sums. Explicitly, we have that:
\be
\begin{split}
S_b&=-\sum_{m=0}^\infty q_2(m) w^{2m+4}\,{}_2F_1[6+2m,6+2m,12+4m,w]+\\
&+\sum_{m=0}^\infty 36 \, q_1(m)\, w^{2m+4}\,{}_2F_1[6+2m,6+2m,12+4m,w]+\\
&+\sum_{m=0}^\infty 288\, q_0(m)\, w^{2m+4}\,{}_2F_1[6+2m,6+2m,12+4m,w]  \,.\\
\end{split}
\ee
The summation in the first line can be readily performed as above. To evaluate the other two sums, it suffices to notice that the coefficients $q_1(m),q_0(m)$ can be expressed as linear combinations of certain $p[a,b,m]$. Explicitly, 
\be
\begin{split}
q_1(m)&={1\over 28800}\left(\frac{15}{28} p[2,4,m]-\frac{1}{2}p[3,3,m]\right),\\
q_0(m)&={1\over 28800}\left(-\frac{5}{84}p[2,4,m]+\frac{5}{252}p[1,5,m]+\frac{1}{24}p[3,3,m]\right)\,.\\
\end{split}
\ee
It then becomes straightforward to perform the relevant summations, leading to:
\be\label{sumb}
\begin{split}
S_b&=\frac{1}{560} \, \left\{-\frac{49}{360} \left(w^2{}_2F_1[3,3,6,w]\right)^2 +\frac{1}{24}  \left(w{}_2F_1[2,2,4,w]\right)\,\left(w^3{}_2F_1[4,4,8,w]\right)+\right. \\
&\qquad\qquad\qquad\qquad\qquad\qquad\qquad\qquad +\left.\frac{1}{9}\,{}_2F_1[1,1,2,w]\, \left(w^4{}_2F_1[5,5,10,w]\right)\right\}\\
\end{split}
\ee
Finally, for $S_c$ we can use the expressions found above relating $q_0(m)$ to certain $p[a,b,m]$, to write:
\be
\begin{split}
S_c&=\frac{1}{400}\left\{  \frac{1}{6} \left(w^2{}_2F_1[3,3,6,w]\right)^2 -\frac{5}{21} \left(w{}_2F_1[2,2,4,w]\right)\,\left(w^3{}_2F_1[4,4,8,w]\right)+\right. \\
&\left. \qquad\qquad\qquad\qquad\qquad\qquad\qquad\qquad +\frac{5}{63} \,{}_2F_1[1,1,2,w]\, \left(w^4{}_2F_1[5,5,10,w]\right) \right\}\,.
\end{split}
\ee
Combining the above we can finally write
\be\label{resTfinalb}
\begin{split}
&\left. G(z,\zbar)\right|_{\mu^2}\simeq_{z\rightarrow 1}[(1-z)(1-\zbar)]^{-\Delta+2}(1-z)^4 \frac{\Delta}{\Delta-2}   \left\{  \Delta^2\frac{1}{28800} \, _2F_1(3,3;6;1-z){}^2+\right. \\
&\left. +\frac{\Delta}{560} \left(-\frac{49}{360} \, _2F_1(3,3;6;1-z){}^2+\frac{1}{24} \, _2F_1(2,2;4;1-z) \, _2F_1(4,4;8;1-z)+\right.\right.\\
&\left.\left. +\frac{1}{9}\, _2F_1(1,1;2;1-z) \, _2F_1(5,5;10;1-z)\, \right)+ \frac{1}{400} \left(\frac{1}{6} \, _2F_1(3,3;6;1-z){}^2\right.\right.\\
&\left.\left.-\frac{5}{21} \, _2F_1(2,2;4;1-z) \, _2F_1(4,4;8;1-z)+\frac{5}{63} \, _2F_1(1,1;2;1-z) \, _2F_1(5,5;10;1-z)\right)\right\} 
\end{split}
\ee
where we restored $u=(1-z)(1-\zbar)$ and set $v\simeq z$ since we are working in the limit $\zbar\rightarrow 1$.


\section{The case of $d=2$}
\label{app:2d}

In this section, we use the fact that in $d=2$, the vacuum block is known exactly in the heavy-heavy-light-light-limit. By expanding in the light-cone limit, we derive the analogous set of OPE coefficients of lowest-twist double-trace built out of the square of the stress-tensor. Moreover, we can explicitly construct these double-trace operators. Not only does this section serve as an example of how the T-channel sum works in 2d, this calculation also provides an insight that multi-trace mixing is important in the large-$C_T$ order that we are working with. The direct computation also suggests that there is perhaps a direct way to generalize the calculation to higher dimension, paving a way to prove the OPE formula we conjectured for holographic  CFTs in 4d.

\subsection{T-channel sum}
From Sec.~4 of \cite{Fitzpatrick:2014vua}, in the heavy-heavy-light-light limit, the Virasoro vacuum block in the $T$-channel dominates, resumming all multi-trace contributions of the stress tensors:
\be
V_{vac}(u,v)\approx \alpha^{\Delta_L} v^{-\half \Delta_L(1-\alpha)} \left[\frac{1-v}{1-v^\alpha}\right]^{\Delta_L}
\ee where $\alpha \equiv \sqrt{1-\mu}$. Note that $C_T=c/2$. We shall drop the subscript `L' from now on and write $\Delta \equiv \Delta_L$.

 Expanded in power of $\mu$ up to order $\mu^2$, one obtains:
\bea
V_{vac}(u,v)
&\approx&
1+\mu \Delta  \left[\frac{(v+1)  }{4 (v-1)}\log v-\frac{1}{2}\right]
\nonumber\\
&&
+\half \mu^2\left\{
\Delta^2 \left[
\frac{(v+1) \log v}{4 (v-1)}-\frac{1}{2}\right]^2
+
\frac{\Delta}{8}\left[
-4
+\frac{(v+1) }{v-1}\log  v
+\frac{2 v }{(v-1)^2}(\log v)^2
\right]\right\}\nonumber\\
&&
+O(\mu^3).
\eea 
From \cite{Kulaxizi:2018dxo}, we know that we can rewrite this as (with $w=1-v$):
\bea\label{eq:2dVac}
V_{vac}
&\approx&
1+\frac{\mu }{24}\Delta  f_2(v)
+\frac{\mu^2}{1152}\left[
(\Delta -2) \Delta f_2(v)^2
+\frac{12 \Delta}{5}f_1(v) f_3(v)
\right]
\nonumber\\
&&
+O(\mu^3).
\eea We remind the readers that
\be
 f_{a}(v)\equiv (1-v)^a\,{}_2 F_1(a,a,2a,1-v) \,.
\ee
Using Eq.~\ref{eq:hyperid}, in the $\mu^2$ term, we can decompose into the S-channel block by
\be
V_{vac} |_{\mu^2} = \Delta\sum_{s\ge 4,even} a_s^2 \left( \Delta+b_s\right) \times w^{s}{}_2 F_1\left[
s,s,2s,w\right]
\ee with
\bea\label{eq:aandb2d}
a_s^2 &=& \frac{\sqrt{\pi } 2^{-2 s-1} (s-2) }{s^2 (s+2) (s-3) (s-1) }\frac{\Gamma (s+2)}{\Gamma \left(s-\frac{1}{2}\right)} \nonumber\\
b_s&=&\frac{4}{(s-2) (s+1)}\,,
\eea where $s\ge 4$ and is even. These are  analogous to the $P_{4,s}$ for $d=4$ given in Eq.~\ref{Ptot}. 
The OPEs of the lowest few spin are given below
\bea\label{app:2dope}
a_4^2\times (\Delta+b_4)&=& \frac{1}{1152}\left(\Delta+\frac{2}{5}\right)\nonumber\\
a_6^2\times(\Delta+b_6)&=&\frac{1}{51840}\left(\Delta+\frac{1}{7}\right)\nonumber\\
a_8^2 \times(\Delta+b_8)&=& \frac{9}{12812800}\left(\Delta+\frac{2}{27}\right)\,.
\eea

\subsection{OPE coefficients by direct calculation}
\label{sec:app2d}
In this section, we shall explicitly construct the OPE coefficients $a_s$ for the double-trace operators built out of $T^2$. Before we begin, we shall set out some convention. 
Let $C_T\equiv c/2$ in 2d CFT and normalize the two-point function of stress tensor to be $1$. In this normalization
First of all, the OPE of stress tensors is given by\footnote{In this normalization,
\be\label{eq:2dOPE3}
T(y) O(0) = \frac{\Delta_O}{2\sqrt{C_T}}  \frac{1}{y^2} O(0) +\frac{1}{\sqrt{C_T}} \frac{1}{y} \partial O(0)+(TO(0))+\ldots
\ee }
\be \label{eq:2dOPE1}
T(y)T(0)=\frac{1}{y^4}  +\frac{2}{\sqrt{C_T}}\left[\frac{1}{ y^2 }  T(0) +\frac{1}{ 2y } \partial T(0)\right]+(T^2)(0) +
\ldots
(T(\partial T) )(0)
+\half y^2 (T(\partial^2 T))(0)+\ldots
\ee

Let  $\Lambda_s$ be the quasi-primary double-trace operator $(T (\partial^{s-4} T))-\ldots$ where $\ldots$ are  single-trace terms. We will write out these single-trace terms explicitly in a moment. Before doing so, we shall make the following comment: To compute $C_{\OO\OO \Lambda_s}$, one would naively thought that the lowest-trace terms will be irrelevant at leading $C_T$. However, we shall now see explicitly that when we compute $C_{OO \Lambda_4}$  that this is not correct. 

In terms of quasiprimary, the OPE reads (suppressing the partial waves)
\be\label{eq:2dOPE2}
T(y)T(0)=\frac{1}{y^4}  +\frac{2}{\sqrt{C_T}}\frac{1}{ y^2 }  T(0) +\sum_{a\ge 0}y^{2a}C _{TT\Lambda_{4+2a}}\Lambda_{4+2a} \,,
\ee  
where  $\Lambda_{4+2a}$ is the quasi-primary double-trace operator $(T \partial^{2a} T)-\ldots$ of spin $s=4+2a$. 
To work out the explicit relations between the  $\Lambda_{4+2a}$ and $(T(\partial^{2a}T))$. We equate the two expressions for the OPE in Eq.~\ref{eq:2dOPE1} and Eq.~\ref{eq:2dOPE2}. First, let us be explicit about the expansion into the quasiprimary for general OPE. Given any two operator $\OO_1$ and $\OO_2$ with holomorphic dimension $h_1$ and $h_2$, the OPE in $z_{12}\equiv z_1-z_2$ is given by:
\be
\OO_1(z_1)\OO_2(z_2)=\sum_{p}C _{\OO_1 \OO_2 \OO_p}C^{(h_1,h_2;h_p)}(z_{12}, \partial_2) \OO_p(z_2) .
\ee where $\OO_p$ is a quasiprimary appearing in the OPE. In 2d, this is given explicit by (see \cite{Belavin:1984vu})
\be
C^{(h_1,h_2;h_p)}(z_{12}, \partial_2)  = {}_1 F_1\left[
h_1-h_2+h_p
,2h_p,
z_{12} \partial_2
\right]\,.
\ee For e.g., this implies that the whole contribution (including descendants) of $T$ in the $TT$ OPE is
\be
T(z_1)T(z_2) |_T= \frac{2}{\sqrt{C_T}}\frac{1}{z^2_{12}}
\left[
1+ \frac{1}{2} z_{12}\partial T(z_2)
+\frac{3}{20} z_{12}^2 \partial^2 T(z_2)
+\frac{1}{30} z_{12}^3 \partial^3 T(z_2)
+\ldots
\right].
\ee 
This contributes to the order $z_{12}^0$ term in the OPE of $TT$ as 
\be
\frac{2}{\sqrt{C_T}} \times \frac{3}{20} \partial^2_z T(z_2)
\ee The relation between $\Lambda_4$ and $T^2$ then differs by this term
\be
(T^2)(z)
=\frac{2}{\sqrt{C_T}} \times \frac{3}{20} \partial^2 T(z)
+C_{TT\Lambda_0} \Lambda_4(z).
\ee 

The OPE $C_{TT\Lambda_s}$  will be fixed by the convention that $\Lambda_{s}$ has unit norm or that its two-point function is unit normalized. After this is done, we obtain
\be
\Lambda_4 =\frac{1}{\sqrt{\frac{22}{5 C_T}+2}}\left[T^2 -\frac{1}{\sqrt{C_T}} \times \frac{3}{10} \partial^2 T(z)\right].
\ee 
In the large $C_T$ limit, we have
\be
\Lambda_4 \approx \frac{1}{\sqrt{2}}\left[T^2 -\frac{1}{\sqrt{C_T}} \times \frac{3}{10} \partial^2 T(z)\right]+O(1/C_T)\ee 
Here, we see that the single-trace term has a $1/\sqrt{C_T}$ coefficient compared to the coefficient of $T^2$.  We shall see that both terms in the square bracket contributes to $C_{\OO \OO \Lambda_4}$ at the same order in the large $C_T$ limit.

Let us now compute $C_{\OO\OO\Lambda_0}$. The first step involve computing $C_{\OO\OO T^2}$. This requires first defining $T^2$ as the (holomorphic) point-splitting limit:
\be
T^2(w) \equiv \oint_{C_w} \frac{dx}{x-w} T(x)T(w) \,,
\ee  where the contour $C_w$ means a closed-contour circling $w$. This trivially reproduces the definition in Eq.~\ref{eq:2dOPE1}. For convenience we have absorbed the $1/(2\pi i)$ in the definition of $\oint$. 
Then according to the generalized Wick's theorem (see around 6.211 of \cite{DiFrancesco:1997nk}), we have
\be
\OO(z) T^2(w)
= \oint_{C_w} \frac{dx}{x-w} \left\{
[\OO(z) T(x)]_{OPE} T(w)
+T(x) [ \OO(z) T(w)]_{OPE}
\right\}
\ee  where $[\ldots]_{OPE}$ means doing OPE on the two operators in the brackets first before performing the contour integral.  We also need to compute
\be
\OO(z) T(w) =\frac{\Delta_\OO}{2\sqrt{C_T}(z-w)^2} \OO(w)+\frac{\half \Delta_\OO-1}{\sqrt{C_T}(z-w)} \OO'(w)+\ldots
\ee by expanding the operators appearing in the OPE of $T(w) \OO(z)$ when $z$ goes to $w$. 

Collecting all the ingredients,  after some straightforward series expansion and picking up the pole, we have
\be
\OO(z) T^2(w)
=
\frac{\Delta_\OO (\Delta_\OO+4) }{4C_T(z-w)^4}\OO(w)
+\frac{(\frac{\Delta_\OO}{2})^2-1 }{C_T(z-w)^3}\OO'(w)
+\ldots
\ee This yields
\be
C_{\OO\OO T^2}=\frac{\Delta_\OO}{4C_T}(\Delta_\OO+4).
\ee The other term (i.e. $-(3/10)\partial^2 T$) is simpler as we can obtain that from
\be
\langle \OO(\infty) \OO(1) T(z)\rangle = \frac{\Delta_\OO}{2\sqrt{C_T}}(z-1)^{-2}=\frac{\Delta_\OO}{2\sqrt{C_T}}
\left[
1+2  z+3  z^2+\ldots
\right]
\ee and comparing with the expansion $T(z)=T(0)+z T'(0)+\frac{1}{2} z^2 T''(0)$, we obtain
\be
C_{\OO\OO \partial^2 T}= \frac{3\Delta_\OO}{\sqrt{C_T}}.
\ee
All in all
\be
C_{\OO\OO T^2}-\frac{3}{10\sqrt{C_T}} C_{\OO\OO \partial^2 T}
=\frac{\Delta_\OO}{4C_T}\left(\Delta_\OO+\frac{2}{5}\right).
\ee We note that in the $\Delta_\OO^2$ term, only $T^2$ contributes while for the $\Delta_\OO$ term, {\it both} the $T^2$ and the $\partial^2 T$ terms contribute at the same order.
We can  then divide by the norm of this state and appropriately normalize and take large $N$ limit to obtain the result from previous section.
The result is
\be
C_{\OO\OO \Lambda_4}= \frac{\Delta_\OO\left(\Delta_\OO+\frac{2}{5}\right)}{4C_T \sqrt{2+\frac{22}{5C_T}}}
\ee which in the large $C_T$ limit becomes
\be\label{eq:2dL4}
C_{\OO_H \OO_H \Lambda_4}
C_{\OO_L \OO_L \Lambda_4}
=\frac{\Delta_H^2}{32C_T^2 } \Delta(\Delta+2/5)
=\frac{\mu^2 \Delta}{1152} \left(\Delta+\frac{2}{5}\right)
\ee where $\Delta_H\equiv C_T \mu/6$ is the dimension of the heavy operator $\OO_H$ while $\Delta$ is the dimension of the light operator $\OO_L$.

Here, we make an important observation: Since we are after $\mu^2$ term, in each OPE coefficient $C_{\OO\OO\Lambda_4}$, we are after $1/C_T$ term. The $C_{\OO\OO T^2}$ term is of order $1/C_T$ as can be roughly estimated from the square of $C_{\OO\OO T}\sim 1/\sqrt{C_T}$. On the other hand, the $C_{\OO\OO\partial^2 T}$ is of order $1/\sqrt{C_T}$ but it comes with a prefactor $1/\sqrt{C_T}$ in the construction of $\Lambda_4$. Therefore, both the double-trace and the single trace terms contribute at order $1/C_T$ in the OPE $C_{\OO\OO \Lambda_4}$.

Similar considerations in the large $C_T$ limit gives
\be
\Lambda_6= \sqrt{\frac{63 }{280 }}
\left[
\half (T'' T)
-\frac{1}{84\sqrt{C_T}}  \partial^4 T
-\frac{5}{36} \partial^2 \Lambda_4
\right]=\sqrt{\frac{63 }{280}}
\left[
\frac{2}{9} (T'' T)
-\frac{5}{18} (T' T')
+\frac{5}{756 \sqrt{C_T}}  T''''
\right] \,,
\ee 
where in the large $C_T$ limit
\be\label{eq:2dL6}
C_{\OO_H \OO_H \Lambda_6}
C_{\OO_L \OO_L \Lambda_6} \approx
  \frac{\mu^2\Delta }{ 51840}\left(\Delta+\frac{1}{7}\right).
\ee 
Finally, for $\Lambda_8$, we have
\be
\Lambda_8=\sqrt{\frac{143}{350}}\left[
\frac{1}{24} (T'' T)
-\frac{1}{4320 \sqrt{C_T}} \partial^6 T
-\frac{7\sqrt{2}}{1584}  \partial^4 \Lambda_4
-\frac{7}{39}\sqrt{\frac{5}{2}} \partial^2 \Lambda_6
\right]
\ee which yields
\be\label{eq:2dL8}
C_{\OO_H \OO_H \Lambda_9}
C_{\OO_L \OO_L \Lambda_8} \approx
\mu^2 \Delta \frac{9}{12812800} ( \Delta+2/27).
\ee Note that the explicitly constructed OPEs in Eq.~(\ref{eq:2dL4}), Eq.~(\ref{eq:2dL6}) and (\ref {eq:2dL8}) agree with Eq.~\ref{app:2dope} which were obtained  from the  Virasoro vacuum block when expanded in the T-channel.

It would be interesting to compute similar OPE coefficients for $d>2$, though there might be a few subtleties and difficulties. The first issue is the definition of $T^2$ in higher-dimension. In 2d, due to holomorphic factorization, we could define $T^2$ nicely through the contour-integral representation above. In higher-dimensional, presumably a carefully constructed point-splitting procedure should yield similar definition. Secondly, due to the tensor-structures of OPE and spinning operator, the generalization of the above computations might be messy. It could be that for the lowest-twist double-trace stress-tensor operator, things simplify and presumably we could focus on light-cone OPE and define $T^2$ appropriately and compute the relevant OPE coefficient. Lastly, one might apply the technology of ``conglomeration'' in \cite{Fitzpatrick:2011dm} to compute this OPE directly in higher-dimensions.

\bibliographystyle{utphys}
\bibliography{muSqDraft}

\end{document}